\documentclass[superscriptaddress,twocolumn,aps,prd,longbibliography]{revtex4-1}

\usepackage[caption=false]{subfig}
\usepackage{lipsum}  
\usepackage{mathtools}
\usepackage{lineno}
\usepackage{hyperref}
\usepackage{graphicx}
\usepackage[normalem]{ulem}
\graphicspath{{./figs/}}
\usepackage{amsmath,amssymb}
\hypersetup{
colorlinks = true,
urlcolor   = blue,
citecolor  = blue,
linkcolor = blue
}
\modulolinenumbers[5]

\renewcommand{\Re}{\mathrm{Re}}
\renewcommand{\vec}[1]{\boldsymbol{#1}}

\usepackage{xcolor}

\newcommand{\av}[1]{\left\langle{#1}\right\rangle}

\begin{document}

\preprint{APS/123-QED}

\title{Velocity alignment explains Lagrangian irreversibility in turbulence}

\author{Ron Shnapp}
\affiliation{Mechanical Engineering Department, Ben Gurion University of the Negev, Beer Sheva, Israel}

\begin{abstract}
	Lagrangian particles in turbulence separate away from each other faster in the backward in time direction as compared to forward in time. In this work, we show that time irreversibility is kinematically rooted in the fact that, when viewed backward in time, the alignment of particles' relative velocities is better than when viewed forward in time.  
\end{abstract}

\pacs{pacs1}
\maketitle



In three-dimensional turbulence, kinetic energy produced at large scales is dissipated into heat at smaller scales. 
This process demands that there is a physical mechanism which transfers kinetic energy from one scale to another.
In turbulence, scale-to-scale energy transfer occurs locally and in both directions, downscale and upscale~\cite{Eyink1995, Borue1998, musacchio2017, Alexakis2018}, yet on a macroscopic level the mean flux is in the downscale direction; this process is called the turbulence cascade.  
The cascade is materialized in the evolution of turbulent flows in two ways. First, to allow the spectral transfer of kinetic energy to occur more in one direction than the other the flow must organize spatially in ways that support this behavior~\cite{Novikov1974, VelaMartin2021, Iacobello2023, Park2025}. The negative skewness of relative velocities associated with vortex stretching~\cite{Batchelor1947} and Kolmogorov's four-fifths law~\cite{Kolmogorov1991} exemplify this fact. Second, the direction of the energy flux from large to small scales also determines the direction of the arrow of time, leading to time irreversibility in turbulence. This dual role of the energy flux thus links space (namely spatial flow structure) and time in turbulence. 

\begin{figure*}
	\centering
	\includegraphics[width=0.94\textwidth]{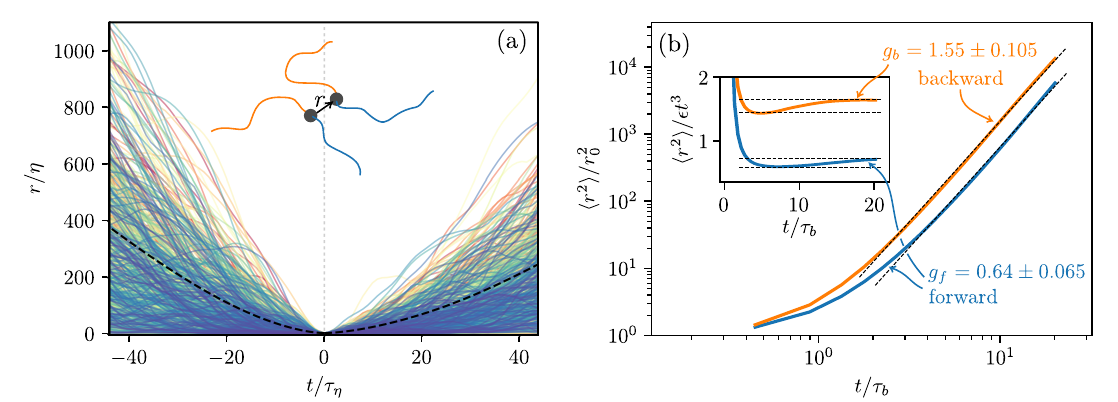}
	\caption{(a) Trajectories of the distance between pairs of Lagrangian particles, backwards and forward in time. Data shown for 1000. A conceptual sketch is shown for two particles separating forward and backward in time, and a definition of the $\vec r$ vector. 
	(b) Pair dispersion either forward of backwards in time. Black lines show Richardson's scaling with estimates of $g$ calculated through the normalized curves shown in the inset.}
	\label{fig:pd}
\end{figure*}

Time irreversibility appear in the Navier Stokes equations only through the viscous term. This fact might lead to the notion that finite energy dissipation is the only source of irreversibility in turbulence. 
However, the fluid motions that support the cascade in the inertial range are to a good approximation inviscid. Therefore, turbulence cascade and the flow organization above the dissipation range are time irreversible not because of the effects of viscosity and energy dissipation.
Accordingly,  Vela-Mart\'{i}n and Jim\'enez~\cite{VelaMartin2021} suggest that there are two kinds of irreversibility in turbulence: one kind is due to energy dissipation, and the second kind is probabilistic in nature. This probabilistic irreversibility is the one which directs the inertial range flow structure towards dynamics that support the downscale cascade~\cite{Novikov1974}, and this process is not reflected in the Navier Stokes equations~\cite{VelaMartin2021}. Indeed, \textit{"The equations of motion alone fix all the solutions but do not choose a distribution from among them}", as Krichnan noted~\cite{Kraichnan1958}.


Understanding how time irreversibility is manifested dynamically could be studied through the motion of fluid tracer particles, so-called Lagrangian irreversibility~\cite{Xu2015}. One aspect of Lagrangian irreversibility is expressed in the power fluctuations of Lagrangian particles that exhibit time asymmetry, so-called flight-crash events~\cite{Xu2014,Grafke2015,Pumir2016,Cencini2017}. 
Interestingly, flight-crash events are not well correlated with local energy dissipation~\cite{DePietro2018}, which echoes the distinction of irreversibility suggested by Vela-Mart\'{i}n and Jim\'enez~\cite{VelaMartin2021}. This type of Lagrangian irreversibility concerns velocity of single particles and thus does not reflect the spatial organization of the flow. Another kind of Lagrangian irreversibility concerns the motion of Lagrangian particles relative to one another. If Lagrangian particles are released locally in a turbulent flow, Richardson's law predicts that the mean squared distance between them grows with time as~\cite{Richardson1926,Salazar2009}
\begin{equation}
\av{r^2} = g_f\epsilon \, t^3
\label{eq:Richardson}
\end{equation}
where $g_f$ is the Richardson constant and $\epsilon$ is the mean kinetic energy dissipation rate; $\av{\cdot}$ denoted an average over Lagrangian particles at fixed times, $t$ after they have been marked~\cite{Monin1972}. When observing this process backward in time the same scaling holds, however the Richardson constant becomes $g_b$ which is more than twice the value of $g_f$~\cite{Sawford2005,Berg2006,Jucha2014,Cheminet2022,Gallon2024}; hence the backward in time separation is much faster than the forward in time, which manifests time irreversibility. Jucha et al.~\cite{Jucha2014} and Drivas~\cite{Drivas2018} suggested an explanation for this phenomenon based on the so-called Lagrangian analogue of the four-fifths law presented in~\cite{Mann1999, Falkovich2001}. However, this law holds only at fixed times, so the explanation applies only the short time behavior of the separation process, while eq.~\eqref{eq:Richardson} and its backward in time analogue apply at longer times, in the inertial range. In other words, the explanation of Refs.~\cite{Jucha2014, Drivas2018} relates to dissipation irreversibility, while an explanation for the probabilistic irreversibility is still missing.

Establishing the ways time irreversibility manifests itself in the flow structure in a dynamic sense requires an understanding of the processes associated with time reversal symmetry breaking across the inertial range. Here, we would like to examine this from the perspective of kinematics. 
The rate of separation of two Lagrangian particles can be written as~\cite{Shnapp2023}
\begin{equation}
\frac{dr}{dt} = |\Delta \vec{v}|\,\cos(\theta)
\label{eq:sep_decomposition}
\end{equation}
where $\Delta \vec{v}$ is the vector of relative velocity between the two particles, and
\begin{equation}
\cos(\theta) = \frac{\vec r \cdot \Delta \vec v}{r\,|\Delta\vec v|}
\end{equation}
is the projection of the particles' relative velocity on their separation vector, $\vec r$ (where $r=|\vec r|$). The angle $\theta$ is the angle between the two vectors, so-called pair dispersion angle, that was associated recently with a universal behavior in turbulence~\cite{Shnapp2023}. From eq.~\eqref{eq:sep_decomposition}, the separation is determined by the interaction of two components: the magnitude of the relative velocity and the alignment between the separation and relative velocity vectors. Thus, time irreversibility can be explained by two phenomena: 1) backward in time, particles typically "choose" paths with higher relative velocities; 2) backward in time, $\vec r$ and $\Delta \vec v$ are typically more aligned, so $\theta$ is typically lower and $dr/dt$ higher. We therefore ask how these two components behave forward and backward in time, and how these behaviors determine the Lagrangian time irreversibility in turbulence?
Below we utilize direct numerical simulation (DNS) results in homogeneous isotropic turbulence and show that the faster separation backward in time occurs because of better alignment in the backward as compared to the forward separation process. 

To explore the source of Lagrangian time irreversibility we utilize a numerical data set of trajectories, integrated from a DNS of the Navier Stokes equations. The DNS was taken from the Johns Hopkins Turbulence Database, where a homogeneous isotropic turbulence at Reynolds number $\Re_\lambda\approx 433$ was forced numerically~\cite{Li2008, Yu2012}. Since its publication, this database was used in numerous studies to uncover fundamental questions regarding turbulent flows. Full details on this openly available resource are given in~\cite{Li2008, Yu2012}, so we shall not repeat this information here for brevity.
Trajectories were integrated in the numerical domain with time steps of $0.2\tau_\eta$ ($\tau_\eta =(\nu/\epsilon)^{1/2}$ is the Kolmogorov timescale, and $\nu$ and $\epsilon$ are the kinematic viscosity and mean rate of turbulent kinetic energy dissipation).

For this work, we integrated 36,903 pairs of trajectories forward in time and 36,903 pairs of trajectories backward in time. The initial positions and times were chosen randomly for each pair across the whole numerical domain $\in[0,\,2\pi^3]$. The initial separation distance for all pairs was kept the same, $r_0 = 3.3\eta$ ($\eta=(\nu^3\epsilon)^{1/4}$ is the Kolmogorov scale). As separation processes are known to strongly depend on the initial conditions~\cite{Shnapp2018, Tan2022}, the choice of fixed initial distance helps to focus the analysis on the time irreversibility rather than other important effects. Furthermore, we chose the initial separation to be $3.3\eta$ as for these values eq.~\eqref{eq:Richardson} scaling holds to good approximation (for other initial distances the scaling is different)~\cite{Ott2000, Elsinga2022, Tan2022, Shnapp2023}. In the analysis, time is normalized by the eddy turnover timescale, $\tau_b=(r_0^2/\epsilon)^{1/3}$, during which the separation process is in the so-called ballistic regime~\cite{Batchelor1952,Bourgoin2006}. Trajectories were integrated for a duration of one Lagrangian integral timescale, equivalent to approximately $20.1\tau_b$, which helps to focus our analysis on the inertial range.



Lagrangian time irreversibility is demonstrated qualitatively for our data set in Fig.~\ref{fig:pd}a, where the time evolutions of the distances between particles is shown as a function of time for 2000 particle pairs. The separation is also analyzed quantitatively in Fig.~\ref{fig:pd}b, in which the mean-squared separation, $\av{r^2}$, is calculated and compared to eq.~\eqref{eq:Richardson} for both forward and backward in time directions. PDFs of the separation are also shown in the supplementary information. While for the forward case the Richardson constant is $g_f = 0.64\pm0.065$, for the backward case it is $g_b = 1.55\pm0.105$ (Fig.~\ref{fig:pd}b inset). Thus, the mean separation in our dataset is higher by a factor of approximately $\sqrt{g_b/g_f}\approx1.55$ in the backward as compared to the forward in time case.

\begin{figure*}
	\centering
	\includegraphics[width=0.94\textwidth]{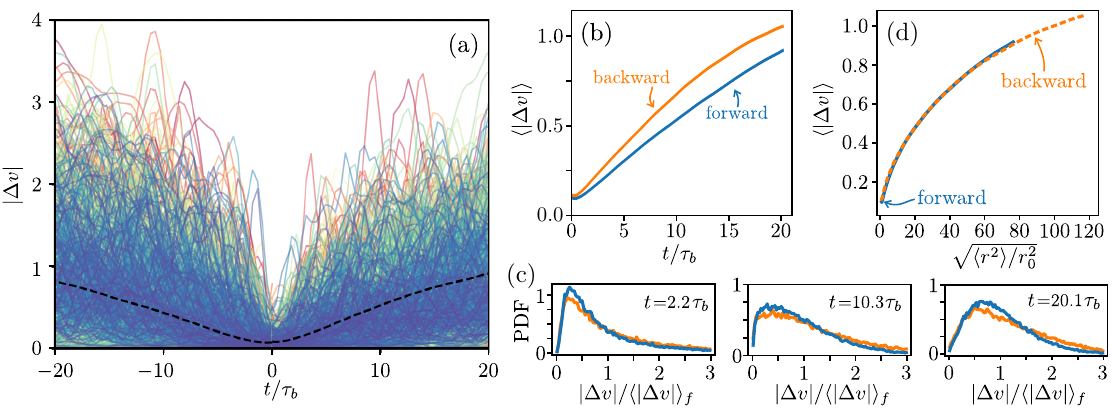}
	\caption{
		(a) Trajectories of the absolute value of the relative velocity between pairs of Lagrangian particles, $|\Delta \vec v|$, backwards and forward in time. Black dashed lines show the median of the distributions.
		(b) Mean relative velocity magnitude plotted as a function of time normalized using the eddy turnover time.
		(c) PDFs of $| \Delta \vec v |$ shown at three time instants, and for the forward and backward in time cases. The distributions are normalized by the mean of $\langle |\Delta v |\rangle$ for the forward direction at each time.
		(d) Mean relative velocity magnitude (same as b), plotted as a function of separation scales, $\sqrt{r^2/r_0}$.}
	\label{fig:rel_valocity}
\end{figure*}


The magnitude of the relative velocity is characterized in Fig.~\ref{fig:rel_valocity}. Trajectories of $|\Delta\vec v|$ are shown in Fig.~\ref{fig:rel_valocity}a for both time directions for 1500 pairs of particles along with the median of the distributions. The mean of $\av{|\Delta \vec v|}$ is shown as a function of time Fig.~\ref{fig:rel_valocity}b and the PDFs of $|\Delta\vec v|$ for three fixed time values are shown in Fig.~\ref{fig:rel_valocity}c. The data show that at each moment in time $|\Delta \vec v|$ typically attains higher values for the backward as compared ot the forward in time direction as expected. And yet, the time asymmetry of $|\Delta \vec v|$ is not as strong as it was for $|\vec r|$; for example, the ratio between the averages of the two cases is approximately $\langle\,\av{|\Delta\vec v|}_b / \av{|\Delta\vec v|}_f \, \rangle \approx 1.24$, which is lower than $\sqrt{g_b/g_f}$.

To understand whether particle pairs sample paths of different relative velocity distributions backward and forward in time requires a consideration of how relative velocities change not with time, but with scale. Indeed, relative velocities in turbulence are well known to grow with the separation distance (e.g. in Kolmogorov scaling and neglecting intermittency $\Delta\vec v\sim r^{1/3}$~\cite{Frisch1995}), and therefore, the mere fact that $g_b>g_f$ is sufficient to cause higher $|\Delta \vec v|$ when probing the particles at fixed times. 
This question is answered in Fig.~\ref{fig:rel_valocity}d, which shows the mean relative velocity $\av{|\Delta \vec v|}$, plotted at fixed scale, namely, as a function of $\sqrt{\av{r^2}/r_0^2}$. Unlike the fixed time observations, here in the forward and backward in time directions the mean relative velocities are the nearly identical. Therefore, the higher values of $|\Delta\vec v|$ observed at fixed times can be explained only by the fact that separation grows quicker in the backward direction. In other words, Fig.~\ref{fig:rel_valocity}d demonstrates that particles do not sample paths with higher $|\Delta \vec v|$ backward in time as compared to forward in time, and thus that $|\Delta \vec v|$ is not the cause of Lagrangian irreversibility.


\begin{figure*}
	\centering
	\includegraphics[width=0.94\textwidth]{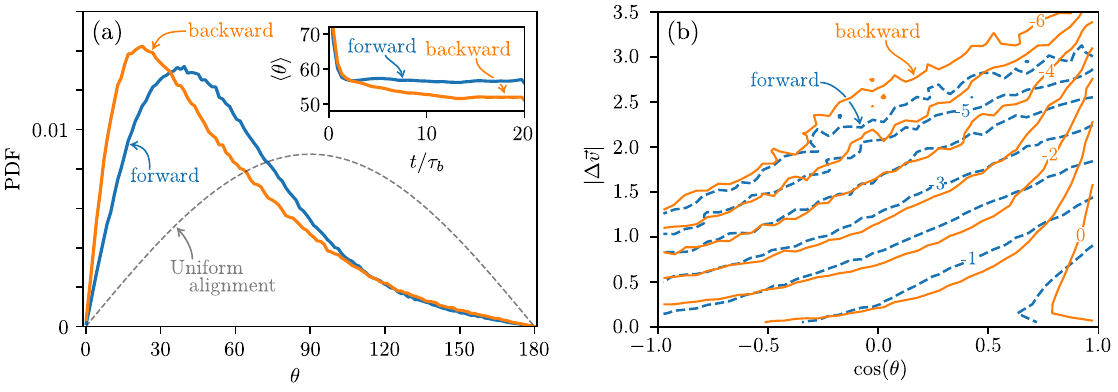}
	\caption{
		(a) PDFs of $\theta$, time averaged in the inertial range for $\tau>6\tau_b$. The inset shows the mean as a function of time. The dashed line shows a reference uniform distribution case.
		(b) Joint PDFs of $|\Delta \vec v|$ and the alignment projection, $\cos(\theta)$. Data shown as isoprobability contours for constant  $\log(P)$ values for the forward (blue dashed) and backward (orange solid) in time cases.}
	\label{fig:angles}
\end{figure*}

%


The second aspect of the separation is the alignment between the relative velocity and the separation vectors. The projection of $\vec r$ on $\Delta\vec v$ ($\cos(\theta)$) is characterized in Fig.~\ref{fig:angles}a through the PDF and the mean value of $\theta$; the projection itself, $\cos(\theta)$, is characterized through its PDF in the supplementary materials. As was recently shown~\cite{Shnapp2023}, $\theta$ has time invariant statistics in the inertial range of turbulence in the forward in time direction. Unlike the separation distance and the relative velocities, the alignment between the vectors is scale-independent, leading to this time invariant behavior. The time invariance is reflected in the mean value, $\av{\theta}$, shown in the inset of Fig.~\ref{fig:angles}a as a function of time. Interestingly, while in the forward time direction $\av{\theta}$ is indeed constant in the inertial range (and equal to $\av{\alpha}=56.6^\circ \pm 1.1^\circ$, in agreement with~\cite{Shnapp2023}), in the backward direction the mean value $\av{\theta}$ slowly decreases in the inertial range, reaching approximately $\av{\theta}\approx50.6^\circ$ at the largest time available $t=20.1\tau_b$. 
The PDFs' evolution across time is also shown in the supplementary information.

The PDFs of the angel $\theta$, averaged in time across the inertial range ($6\tau_b<t<T_L$), is shown in Fig.~\ref{fig:angles}a for the forward and backward in time cases. As expected, the distributions are strongly biased towards the mutually aligned orientations, $\theta\ll90^\circ$. Moreover, the alignment bias is significantly stronger in the backward as compared to the forward in time direction. In quantitative terms, the most probable value of $\theta$ is approximately $22.6^\circ$ in the backward direction as compared to $39.5^\circ$ in the forward direction. 
It is interesting to note that the time asymmetry is much more pronounced in the mutually aligned as compared to the anti-aligned states, namely, the PDFs are significantly different for $\theta<90^\circ$ while they are nearly identical for $\theta>90^\circ$. Therefore, the time asymmetry is much more pronounced in the most probable values $\theta$ distributions as compared to the means. 
All in all, as alignment between $\vec r$ and $\Delta \vec v$ is a necessary condition for the separation distance between particles to grow and as better alignments lead to faster separation (smaller $\theta<90^\circ$ leads to higher $\cos(\theta)$ which increases $dr/dt$ through eq.~\eqref{eq:sep_decomposition}), the results in Fig.~\ref{fig:angles}a suggest that Lagrangian irreversibility can be explained by the smaller of values of $\theta$ observed for the backward direction. More over, because $|\Delta \vec v|$ is not sensitive to the direction of time at fixed scales (Fig.~\ref{fig:rel_valocity}d), better alignments alone explain Lagrangian irreversibility with respect to the decomposition proposed in eq.~\eqref{eq:sep_decomposition}.

Before concluding, it is important to note that the alignment and relative velocities are strongly correlated with one another. In trajectories for which the separation and relative vectors are more aligned (lower $\theta$) the separation process is more efficient; thus, as the characteristic relative velocities are higher for larger separations (Fig.~\ref{fig:rel_valocity}d), we expect that for pairs with higher $\cos(\theta)$ values the relative velocities should also acquire higher values. This correlation is shown to hold true for both the forward and backward in time cases, as shown through the joint PDFs of $|\Delta\vec v|$ and $\cos(\theta)$ in Fig.~\ref{fig:angles}b. Interestingly, the higher separation rates for the backward as compared with the forward case also cause the PDFs to qualitatively differ for both cases. Specifically, isoprobability contours bend upwards with $\cos(\theta)$ for the backward case, while they are approximately straight for the forward case. Therefore, the increase of the relative velocities with $\cos(\theta)$ tends to be much stronger in the backward as compared to the forward in time direction. For the backward separation case, both the relative velocities are higher and the alignment between the separation and relative velocity vectors is better, eventually leading to significantly faster separation rates, all caused by the better alignment between $\vec r$ and $\Delta \vec v$.


To conclude, in turbulent flows the fact that energy is flowing from large to small scales sets the direction of the arrow of time. In the Lagrangian framework, time reversal symmetry is broken as pairs of particles separate significantly faster backward in time as compared to forward in time. As this Lagrangian irreversibility is maintained throughout the inertial range, we understand this phenomenon as the manifestation of the flow spatially organizing in ways that support the downward cascade. This work shows that the flow organization is associated with better alignment properties between the relative velocity and separation vector, and that this alone explains the faster backward in time separation of Lagrangian particles. This realization could support future efforts of constructing statistical models of turbulence. For example, similar to the fact that in kinetic theory minimization of entropy is utilized to form the Maxwell distribution, ideas about separation alignment might help in constructing relative velocity distributions in turbulent flows.

\begin{acknowledgements}
	
\noindent I would like to thank Stefano Brizzolara, Marius M. Neamtu-Halic, Alessandro Gambino, Markus Holzner, and Yoav Green for their comments and discussions. I acknowledge financial support from ISF grants 1244/24 and 2586/24 and from the Alon Scholarship.
	
\end{acknowledgements}

\bibliography{bib}

\end{document}